\documentclass{ieeeaccess}
\usepackage{academicons}
\usepackage{algorithm}
\usepackage{algpseudocode}
\usepackage{amsmath, amssymb, amsfonts}
\usepackage{amsthm, enumitem}
\usepackage{array}
\usepackage{booktabs}
\usepackage{caption}
\usepackage{color}
\usepackage{comment}
\usepackage{float}
\usepackage[T1]{fontenc}
\usepackage{graphicx}
\usepackage{hyperref}
\usepackage[utf8]{inputenc}
\usepackage{lipsum}
\usepackage{longtable}
\usepackage{multirow}
\usepackage{stfloats}
\usepackage{subcaption}
\usepackage{mathptmx}
\usepackage{textcomp}
\usepackage[flushleft]{threeparttable}
\usepackage{url}
\usepackage{wrapfig}
\usepackage[numbers]{natbib}

\def\BibTeX{{\rm B\kern-.05em{\sc i\kern-.025em b}\kern-.08em
T\kern-.1667em\lower.7ex\hbox{E}\kern-.125emX}}
\def\orcid#1{\kern .08em\href{https://orcid.org/#1}{\includegraphics[keepaspectratio,width=0.7em]{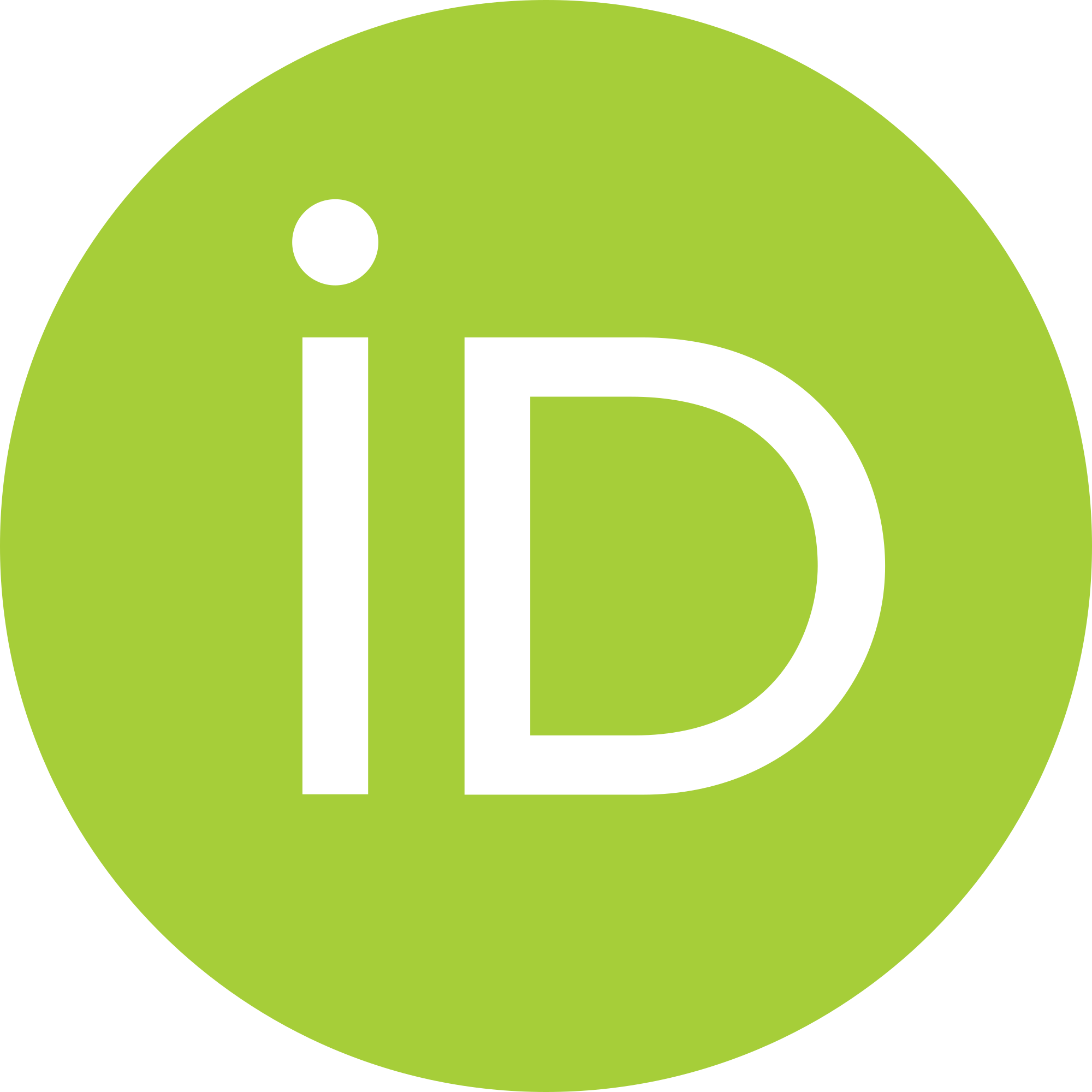}}}

\begin{document}
\history{Date of publication xxxx 00, 0000, date of current version xxxx 00, 0000.}
\doi{10.1109/ACCESS.2023.0322000}

 \title{Securing 5G and Beyond-Enabled UAV Networks: Resilience Through Multiagent Learning and Transformers Detection}

\author{
{Joseanne Viana}\authorrefmark{1, 2}\textsuperscript{\textsection} \orcid{0000-0002-4191-3127} \and \IEEEmembership{IEEE Member} ,
{Hamed Farkhari}\authorrefmark{3}\textsuperscript{\textsection} \orcid{0000-0002-2620-260X} \and \IEEEmembership{IEEE Member} ,
{Victor P Gil Jimenez}\authorrefmark{2} \orcid{0000-0001-7029-1710} \and \IEEEmembership{IEEE Senior Member}}

\address[1]{Tyndall National Institute, Ireland; }
\address[2]{UC3M - Universidad Carlos III de Madrid, Madrid, Spain;} 
\address[3]{ISCTE – Instituto Universitário de Lisboa, Av. das Forças Armadas, 1649-026 Lisbon, Portugal;}

\address {hamed\_farkhari@iscte-iul.pt (H.F); joseanne.viana@tyndall.ie (J.V); vgil@ing.uc3m.es (V.G);}

\corresp{Corresponding author: Joseanne Viana (joseanne.viana@tyndall.ie)}

\begin{abstract}
Achieving resilience remains a significant challenge for Unmanned Aerial Vehicle (UAV) communications in 5G and 6G networks. Although UAVs benefit from superior positioning capabilities, rate optimization techniques, and extensive line-of-sight (LoS) range, these advantages alone cannot guarantee high reliability across diverse UAV use cases. This limitation becomes particularly evident in urban environments, where UAVs face vulnerability to jamming attacks and where LoS connectivity is frequently compromised by buildings and other physical obstructions. This paper introduces DET-FAIR-WINGS ( Detection-Enhanced Transformer Framework for AI-Resilient Wireless Networks in Ground UAV Systems), a novel solution designed to enhance reliability in UAV communications under attacks. Our system leverages multi-agent reinforcement learning (MARL) and transformer-based detection algorithms to identify attack patterns within the network and subsequently select the most appropriate mechanisms to strengthen reliability in authenticated UAV-Base Station links.
The DET-FAIR-WINGS approach integrates both discrete and continuous parameters. Discrete parameters include retransmission attempts, bandwidth partitioning, and notching mechanisms, while continuous parameters encompass beam angles and elevations from both the Base Station (BS) and user devices. The detection part integrates a transformer in the agents to speed up training.  Our findings demonstrate that replacing fixed retransmission counts with AI-integrated flexible approaches in 5G networks significantly reduces latency by optimizing decision-making processes within 5G layers.
The proposed algorithm effectively manages the diverse discrete and continuous parameters available in 5G and beyond to enhance resilience. 
Our results show that DET-FAIR-WINGS achieves faster convergence, reaching high performance in 400 epochs versus 800 for other approaches. The framework approach maintains lower packet loss rates, with 80\% of measurements below 0.2 compared to 70\% for alternatives and delivers lowest latency (mean: 14.61ms, median: 10.62ms) with less variance than Independent Proximal Policy Optimization (IPPO) and  Multi-Agent Proximal Policy Optimization (MAPPO).

\end{abstract}

\IEEEoverridecommandlockouts

% \titlepgskip=-05pt
\setlength{\textfloatsep}{4pt }

\maketitle
\renewcommand\thefootnote{\textsection}
\footnotetext[1]{Collaborative authors with equal contribution}
\renewcommand\thefootnote{\arabic{footnote}}

\section{Introduction}

The widespread adoption of Unmanned Aerial Vehicle (UAV) technology has created unprecedented opportunities across commercial sectors, from delivery services to emergency response operations. As the industry progresses toward 6G communications, ensuring secure and reliable UAV operations presents significant challenges \cite{ref1}. A critical focus in UAV 6G standardization lies in enhancing coverage, reliability, and security across diverse operational conditions.

Industry standards currently target a reliability of $R = 10^{-3}$ for UAV communications, yet this proves insufficient for critical applications. UAVs face considerable challenges in maintaining stable line-of-sight (LoS) connections, particularly in dense urban environments such as New York or Shanghai. The complexity of UAV channel models varies with environmental topology, vehicle speed, and surrounding telecommunications infrastructure. Furthermore, the aerial nature of these platforms makes them susceptible to various jamming threats, including terrestrial, aerial, and static interference sources \cite{Skokowski2024, ref2}.

Contemporary telecommunication systems utilize various recovery and feedback mechanisms to enhance communication reliability across different fading scenarios. While recent innovations in bandwidth partitioning and beamforming, alongside established techniques like adaptive modulation and hybrid automatic repeat request (HARQ) retransmission protocols, offer incremental improvements in link quality metrics, they fall short in ensuring the reliability demanded by critical UAV operations. Key limitations include:

\begin{enumerate}
    \item Adaptive modulation mechanisms struggle with rapidly changing channel conditions between adaptation and transmission phases \cite{VianaJ2022Access}
    \item HARQ protocols, while beneficial, often result in increased retransmission overhead and reduced effective throughput \cite{Tyrovolas2022}
    \item Traditional beamforming approaches, though effective in static networks, encounter significant challenges in high-mobility scenarios \cite{Song2021}
    \item Current positioning and data rate optimization methods prove inadequate for ensuring reliability in dynamic UAV applications \cite{ref2}
\end{enumerate}

These limitations are intensified by the fragmented nature of parameter management in current network architectures. Traditional 5G/6G networks implement communication parameters through segregated protocol stack layers with insufficient inter-layer synchronization mechanisms, creating optimization silos that substantially prevents coordinated adaptation. The hierarchical isolation between physical layer (PHY) modulation and coding schemes (MCS), radio link control (RLC) segmentation parameters, and medium access control (MAC) scheduling algorithms results in compartmentalized optimization frameworks that operate with incomplete system-wide state information.

Beamforming angles are controlled primarily at the Physical (PHY) layer through complex precoding matrices and antenna weighting coefficients, with limited visibility into higher-layer requirements. While the PHY layer is responsible for estimating channel state information (CSI) and computing optimal beam patterns, it does so with incomplete knowledge of application demands or mobility patterns, leading to suboptimal adaptation in dynamic environments.

Bandwidth allocation is determined by the MAC scheduler using scheduling algorithms (e.g., proportional fair, round-robin, or maximum throughput), which operate with minimal awareness of real-time PHY layer conditions or RF interference patterns. Resource Block (RB) assignments and Bandwidth Part (BWP) selections rely on generalized QoS metrics rather than fine-grained channel characteristics. Meanwhile, interference mitigation through notching is handled at the PHY layer’s subcarrier processing level, where specific Resource Elements (REs) or subcarriers are zeroed or power-reduced—often without coordinated awareness of neighboring cell interference or dynamic spectrum usage patterns.

Retransmission counts are configured independently by the MAC layer’s HARQ entity, which determines maximum retransmission attempts, combining methods (Chase Combining or Incremental Redundancy), and timing parameters. These configurations are made with limited feedback from the PHY layer regarding current beamforming effectiveness or bandwidth utilization, potentially leading to suboptimal tradeoffs between reliability and latency. Additionally, the RLC layer introduces another independent retransmission control mechanism through its ARQ process, often operating with different policies and objectives than MAC-layer HARQ, further complicating cross-layer optimization.

\paragraph{Literature Review}

Multi-Agent Proximal Policy Optimization (MAPPO) has emerged as a  algorithm for complex multi-agent reinforcement learning challenges. Fundamental work by Yu et al. \cite{yuPPO2021} demonstrated MAPPO's  performance in cooperative, multi-agent environments, establishing its potential for wireless communications applications. Recent research has expanded MAPPO's applications significantly: Chen et al. \cite{Chen2024} applied it to UAV cooperative combat maneuvering, while Lyu et al. \cite{Lyu2023} demonstrated its effectiveness in complex multi-agent simulations. Building on this foundation, Huang et al. \cite{Huang2022} integrated attention behavior networks to improve feature extraction in high-dimensional state spaces—an innovation particularly valuable for sophisticated inter-agent coordination scenarios.

MAPPO's adaptability has led to diverse applications across wireless communications domains. Cai et al. \cite{Cai2023} optimized spectrum allocation in vehicular networks, Yan et al. \cite{Yan2024} enhanced energy trading in multi-base-station systems, and Peng et al. \cite{Peng2022} improved content caching in integrated satellite-terrestrial networks using hierarchical critic architectures. These applications demonstrate MAPPO's versatility in addressing resource allocation challenges across different network architectures.

For UAV-specific applications, Liu et al. \cite{10021296} proposed an enhanced MAPPO approach with attention mechanism and Beta distribution (AB-MAPPO) for energy-efficient computation offloading in aerial edge networks. Their implementation leverages centralized training with decentralized execution to coordinate multiple UAVs and mobile users, achieving significant improvements in resource utilization and energy efficiency. Similarly, Li et al. \cite{li2023adaptive} demonstrated MAPPO's effectiveness with attention mechanisms for UAV-assisted networks, successfully managing complex hybrid action spaces for communication optimization. Wang et al. \cite{wang2024uav} showed MAPPO's superiority for UAV trajectory design in Internet of Vehicles environments, with energy consumption reductions of 11-19\% compared to alternative multi-agent reinforcement learning methods such as Multi-Agent Deep Deterministic Policy Gradient (MADDPG).

In emergency and specialized UAV deployments, Guan et al. \cite{Guan2023} demonstrated MAPPO's effectiveness in coordinating UAVs for emergency communications in disaster zones, complementing Jiang et al.'s \cite{Jiang2024} findings in underwater applications. This versatility stems from MAPPO's balanced approach to exploration and exploitation in multi-agent contexts, enabled by centralized training with decentralized execution. Kang et al. \cite{Kang2024} developed a transformer-based approach for dynamic avatar task migration in UAV-assisted vehicular Metaverses, showing how strategic pre-migration can significantly reduce service latency in high-mobility environments.

Several research efforts have addressed security challenges in UAV communications. Lv et al. \cite{10279067} developed a multi-agent reinforcement learning framework for UAV swarm communications under jamming attacks, demonstrating effective relay selection and power allocation policies that improve communication resilience while reducing energy consumption. Cai et al. \cite{10506576} introduced a privacy-driven security-aware task scheduling mechanism for Space-Air-Ground Integrated Networks using MAPPO, jointly optimizing delay, energy consumption, and security utility while maintaining different privacy protection levels.

For cellular and edge computing applications, Yin and Yu \cite{9475535} developed a MAPPO framework that jointly optimizes resource allocation and trajectory design in UAV-aided cellular networks, achieving near-optimal performance without centralized control. Zhao et al. \cite{9725258} addressed task offloading in UAV-based mobile edge computing systems, proposing a framework that balances system performance through distributed policy learning. Jia et al. \cite{10720868} introduced a comprehensive framework optimizing satellite edge computing through joint consideration of task queue delays and dynamic loads. Miuccio et al. \cite{10624759} focused specifically on feasible wireless MAC protocol design using multi-agent reinforcement learning, addressing the critical challenge of implementing RL-based approaches with minimal overhead in practical communications systems.

Innovations in MAPPO implementation have improved its scalability for complex environments. The K-MAPPO method \cite{Guan2023MAPPOBasedCU} combines an enhanced K-means algorithm with multiagent reinforcement learning to reduce interaction overhead and improve deployment efficiency—enabling UAVs to operate cooperatively without requiring a central controller. Kang et al. \cite{Kang2023CooperativeUR} formulated UAV resource allocation as a partially observable Markov decision process, where UAVs make cooperative decisions on ground device associations, resource allocation, and task offloading based on local observations. Qin et al. \cite{Qin2024} introduced an effective methodology for decomposing multi-dimensional optimization challenges in Non-Orthogonal Multiple Access (NOMA)-enabled multi-UAV collaborative caching networks, separating caching decisions from trajectory planning and resource allocation to achieve near-optimal performance with reduced computational complexity.

Unlike previous approaches that typically focus on a single network layer or optimization objective, our work represents the first application of MAPPO to joint PHY-MAC layer optimization specifically targeting ultra-reliability in UAV communications. While prior work has demonstrated MAPPO's effectiveness for energy efficiency \cite{10021296}, coverage optimization \cite{Guan2023}, and security \cite{10279067}, none has specifically addressed the stringent reliability requirements ($R = 10^{-6}$) needed for critical UAV applications. Our approach builds upon insights from these diverse applications—combining the hybrid action space handling demonstrated by Liu et al. \cite{10021296} and Li et al. \cite{li2023adaptive}, the adversarial robustness principles from Lv et al. \cite{10279067}, and the implementation feasibility considerations from Miuccio et al. \cite{10624759}. By simultaneously optimizing beamforming, HARQ configuration, bandwidth selection, and interference mitigation, our framework represents a significant advance in achieving ultra-reliable UAV communications for mission-critical applications.

Table \ref{tab:mappo_overview} summarizes the contributions of the main papers in the literature review and highlight the contributions of this paper. 

Rather than relying on traditional network parameter configurations that separately define beamforming alignment angles, retransmission mechanisms, bandwidth allocations, and interference notching based on modulation and scheduling decisions, our work proposes a Multi-Agent Reinforcement Learning (MARL) framework that jointly optimizes these parameters to increase reliability in UAV attacks scenarios. This approach represents a significant shift from conventional network configuration methodologies toward a coordinated, adaptive system that can respond to changing environmental conditions in real-time using Artificial Intelligence (AI) mechanisms.

Our approach integrates several complementary techniques to enhance resilience against jamming attacks:

\begin{itemize}
   \item \textbf{Transformer-based jamming detection:} Leverages attention mechanisms to identify subtle patterns in received signal 
    \item \textbf{Adaptive beamforming optimization:} Dynamically adjusts spatial signal directionality to maintain connectivity while avoiding  unreliable zones
    \item \textbf{Context-aware HARQ configuration:} Modifies  max retransmission  attempts based on detected threat characteristics 
    \item \textbf{Dynamic bandwidth part selection:} Allocates frequency resources to minimize exposure to spectral interference
    \item \textbf{Precision interference notching:} Implements targeted frequency suppression at identified jamming bands while preserving signal integrity
    \item \textbf{Coordinated MAPPO-based response:} Orchestrates multi-agent policy optimization to maximize system resilience against attacks
\end{itemize}

This multi-agent approach enables comprehensive, adaptive decision-making that responds effectively to environmental changes and security threats. Additionally, we enhance the system's resilience by integrating a transformer-based jamming detector operating in inference mode.  This integration improves the framework's ability to detect and respond to jamming attacks while maintaining optimal network performance.

The remainder of this paper is organized as follows: Section \ref{sec:scenario_model_description} presents a detailed scenario and mathematical formulation for the UAV communication. Section \ref{sec:Anti_Jamming} introduces changes in the network that the agents are able to do. Section \ref{sec:mappo_framework} describes the multi-agent system we developed. Section \ref{sec:transformer} describes the transformer integration and how its classification logits enhance the MAPPO framework. Section \ref{sec:results} discusses simulation results. Section \ref{sec:conclusion} provides conclusions and future research directions.

\begin{table*}[ht]
\caption{Overview of MAPPO Applications in Wireless Communications}
\label{tab:mappo_overview}
\centering
\scriptsize % Reduced font size from normal to small
\begin{tabular}{|p{0.6cm}|p{5.0cm}|p{3.0cm}|p{1.3cm}|p{1.2cm}|p{1.5cm}|p{2.3cm}|}
\hline
\textbf{Ref.} & \textbf{MAPPO Innovation} & \textbf{Key Features} & \textbf{Network Layer} & \textbf{Training Approach} & \textbf{Action Space} & \textbf{Optimization Target} \\
\hline
\cite{yuPPO2021} & Original MAPPO algorithm Cooperative Games & Basic algorithm structure & N/A & Centralized & Discrete & Task completion \\
\hline
\cite{Cai2023} & Dynamic spectrum prioritization  Vehicular Networks & Spectrum allocation & MAC & Centralized & Discrete & Spectrum allocation \\
\hline
\cite{Peng2022} & Hierarchical critic architecture Satellite-terrestrial & Content caching & APP & Hierarchical & Discrete & Cache utilization \\
\hline
\cite{10624759} & Feasible MARL implementation with minimal overhead MAC Protocol Design & Protocol learning, reward design & MAC & CTDE & Discrete & Communication efficiency \\
\hline
\cite{Yan2024} & NTP-MAPPO with Nash trading points Base-station Networks & Energy trading & APP & Distributed & Continuous & Energy efficiency \\
\hline
\cite{10720868} & Problem decomposition for scalability Satellite Edge & Task queue management & APP & Decomposed & Continuous & Delay reduction \\
\hline
\cite{Huang2022} & Attention behavior network integration Feature Extraction & Feature selection & N/A & Centralized & High-dimensional & Coordination \\
\hline
\cite{10506576} & Multi-level encryption configuration SAGIN Security & Security, privacy protocols & APP, NET & Multi-objective & Hybrid & Security utility \\
\hline
\cite{Guan2023MAPPOBasedCU} & K-MAPPO with clustering preprocessing UAV Disaster Response & Clustering, deployment & NET & Decentralized & Discrete & Deployment efficiency \\
\hline
\cite{Kang2023CooperativeUR} & Partial observation handling UAV Aerial Computing & Partial observability, task allocation & APP & CTDE & Discrete & Task computation \\
\hline
\cite{Chen2024} & GRU for temporal feature extraction UAV Air Combat & Temporal features, maneuvering & APP & Centralized & Continuous & Maneuvering \\
\hline
\cite{9475535} & Independent value function estimation UAV Cellular & Resource allocation & PHY & Distributed & Continuous & Network performance \\
\hline
\cite{wang2024uav} & Content-aware reward shaping UAV Content Caching & Content placement, trajectory & APP, NET & CTDE & Continuous & Energy consumption \\
\hline
\cite{Guan2023} & Cooperative trajectory optimization UAV Emergency Comms & Trajectory, coverage & NET & Centralized & Hybrid & Coverage \\
\hline
\cite{Kang2024} & Transformer-augmented state encoding UAV Metaverse & Task migration, latency & APP & Centralized & Hybrid & Service latency \\
\hline
\cite{10279067} & Adversarial training against jammers UAV Anti-jamming & Jamming detection, relay selection & PHY & Distributed & Hybrid & Resilience \\
\hline
\cite{9725258} & Cooperative offloading mechanism UAV Edge Computing & Task offloading, resource allocation & APP & Distributed & Hybrid & System balance \\
\hline
\cite{Jiang2024} & Age of Information integrated rewards AUV Swarms & Data collection, AoI & APP, NET & Distributed & Hybrid & Data freshness \\
\hline
\cite{Qin2024} & Decomposition of multidimensional DRL Multi-UAV Caching & Content caching, trajectory & APP, NET & Hierarchical & Hybrid & Resource allocation \\
\hline
\cite{10021296} & AB-MAPPO with modified policy network UAV Edge Computing & Computation offloading & APP & CTDE & Hybrid & Energy efficiency \\
\hline
\cite{li2023adaptive} & Digital twin integration for state prediction UAV Networks & Sensing, communication optimization & PHY, NET & CTDE & Hybrid & Network optimization \\
\hline
\textbf{This work} & \textbf{First MAPPO-based joint PHY-MAC layer optimization for ultra-reliability} \textbf{UAV Reliable Communications} & \textbf{Retransmissions, delay, Resource allocation, Beam Selection} & \textbf{PHY, MAC} & \textbf{CTDE} & \textbf{Hybrid} & \textbf{reliability } \\
\hline
\end{tabular}
\end{table*}

\section{Network Scenario and Model Description}
\label{sec:scenario_model_description}

This research addresses the challenge of ensuring reliability in UAV communications under jamming attacks.

\begin{figure}[htbp]
    \centering
    \includegraphics[width=0.48\textwidth]{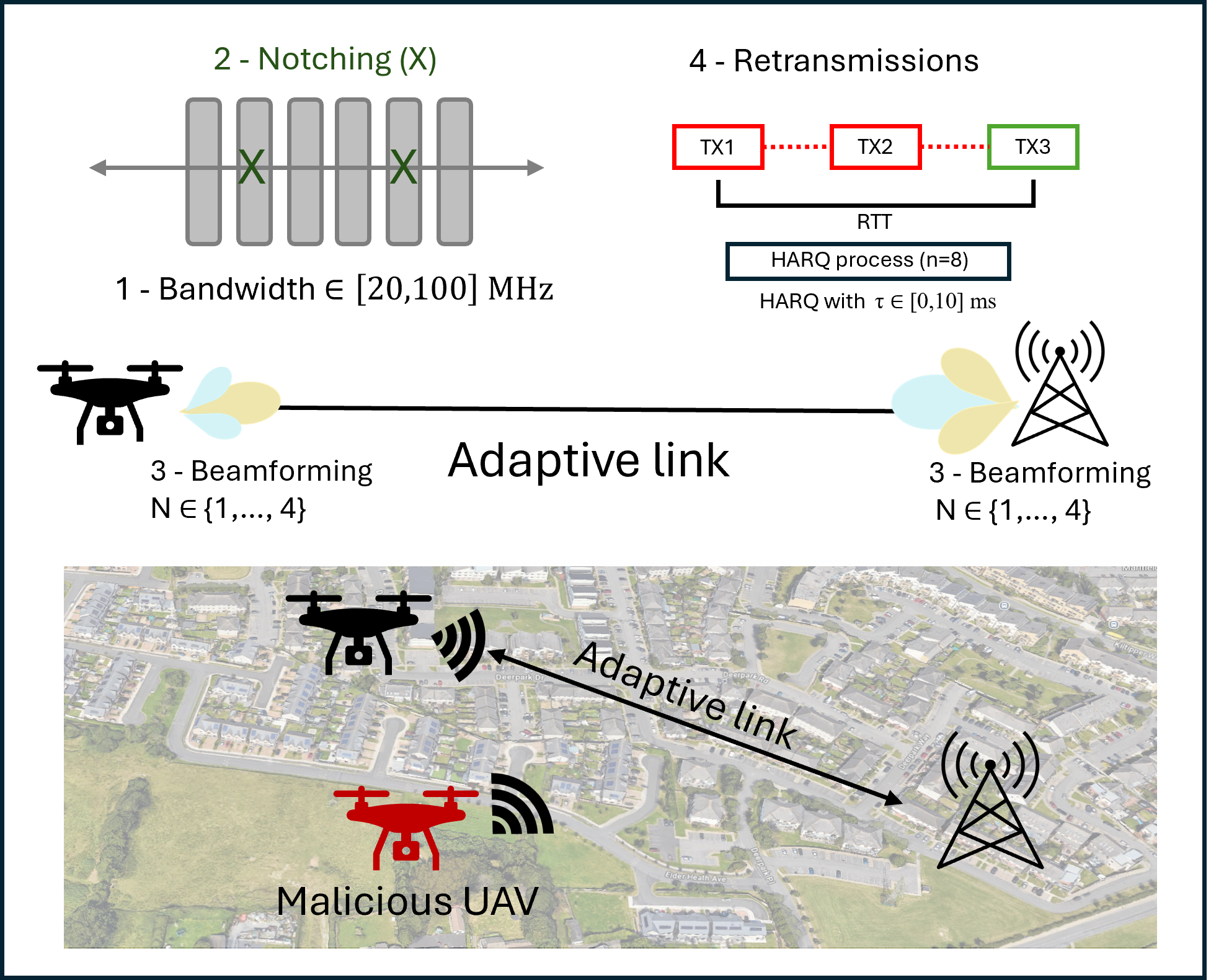} 
    \captionsetup{font=footnotesize}
    \caption{System scenario where a legitimate drone transmits signals to a base station over an adaptive wireless link, with a malicious UAV attempting to interfere with the connection. The diagram highlights four key configurable parameters that can optimize the legitimate link: (1) adjustable bandwidth from 20-100 MHz, (2) frequency notching to mitigate interference, (3) beamforming with configuration options at both endpoints, and (4) retransmission mechanisms using HARQ technique.}
    \label{fig:1}
\end{figure}

As illustrated in figure \ref{fig:1}, our framework integrates four key optimization mechanisms to establish an adaptive link between UAVs and base stations: (1) dynamic bandwidth selection within a 20-100 MHz range, (2) selective notching to mitigate interference at specific frequency bands, (3) optimized beamforming with up to $4$ antenna elements at both the UAV and base station, and (4) intelligent retransmission strategies with HARQ process parameters. The lower portion of the figure demonstrates our system's real-world application, where authenticated UAVs must maintain reliable communications despite the presence of malicious jammers in urban environments.
At the core of our contribution is a MAPPO based solution that determines the optimal combination of these four mechanisms in real-time. Our approach uniquely handles both discrete parameters (bandwidth selection, notching patterns, retransmission counts) and continuous parameters (beamforming angles and elevations) within a unified framework. By simultaneously optimizing across traditionally separate network functions, our solution responds dynamically to jamming attacks, adapting transmission parameters to maintain communication integrity even when malicious UAVs attempt to disrupt the network. This coordinated adaptation pushes reliability toward level required for mission-critical UAV applications, significantly outperforming conventional systems that manage these parameters in isolation.

We propose a scenario consisting of a 5G network deployed in a 1 km$^2$ urban area. The network comprises $N$ antennas, authenticated UAVs, and gNBs, all randomly distributed across the terrain. To accurately model signal propagation in this complex environment, we employ a air-to-ground channel model that incorporates both LoS and NLoS connections. Authenticated UAVs move according to a realistic mobility model, maintaining a variable speed $v(t)$ within the range $[v_{min}, v_{max}]$. The trajectory of each UAV is moving away from the base station or the UAV  static to maintain network connectivity, simulating operations in urban environments.
To create a realistic urban landscape, we randomly distribute buildings of varying sizes and heights across the simulated area. The building parameters follow statistical distributions derived from real urban data, ensuring a representative model of signal obstruction and reflection in a typical urban setting.
For a more detailed explanation of the scenario parameters and simulation methodology, readers are directed to our previous work \cite{Viana2024}, \cite{SyntheticJViana}.

\subsection{Path Loss Model}

The path loss for our scenario is mathematically expressed as:

\begin{equation}
PL_{j}^{\alpha}(d_j, f_c) = 
\begin{cases} 
\gamma(f_c) \cdot d_{j}^{-\alpha} & \text{if LoS}\\
\eta(f_c) \cdot d_{j}^{-\alpha} & \text{if NLoS}
\end{cases}
\label{eq:pathloss}
\end{equation}

where $d_{j}$ represents distance between the UAV and the gNB, $\alpha$ is the path loss exponent, $j$ denotes a specific communication link within the network topology, and $\gamma(f_c)$ and $\eta(f_c)$ are frequency-dependent, environment-specific constants that capture the distinct characteristics of LoS and NLoS propagation, respectively, with $f_c$ representing the center frequency.

\subsection{Fading Model}

To model small-scale fading effects, we utilize the Cluster-Delay-Line (CDL) models as specified in 3GPP standards \cite{3GPP38901, 3GPP36777}. These models provide a comprehensive set of parameters for fading simulation, including:

\begin{itemize}
    \item Power levels of multipath components
    \item Signal delays
    \item Angles of arrival (AoA) and departure (AoD)
    \item Azimuth and zenith spreads (Azimuth Spread of Arrival—ASA, Departure—ASD, Zenith Spread of Arrival—ZSA, and Departure—ZSD) for each cluster
\end{itemize}

Our scenario incorporates both large-scale and small-scale fading effects in the UAV-to-small cell links, providing a realistic representation of the challenging wireless environment.

\subsection{ Extended Received Power Calculation with Multiple Interferers}

In complex urban environments, UAVs often encounter interference not only from intentional jamming but also from multiple surrounding base stations. To accurately model this scenario, we extend our previous analysis to incorporate these additional interference sources, providing a more comprehensive representation of the urban communication landscape.

Our baseline model for the received power at a UAV, in the absence of interference, remains unchanged:

\begin{equation}
P_{uav, j} = P_{tx} + G_{total} - PL_{j}^{\alpha} (d_j, f_c) - S - X_{\sigma}
\label{eq:power}
\end{equation}

where $P_{tx}$ is the transmit power, $G_{total}$ is the combined antenna gain, $PL_j^{\alpha}(d_{j}, f_c)$ represents the path loss, $S$ accounts for small-scale fading (from CDL model), and $X_{\sigma}$ represents log-normal shadowing.

We extend the Signal-to-Interference-plus-Noise Ratio (SINR) formulation to account for both multiple interfering base stations and intentional jamming.

To study network resilience, we introduce a set of adversarial nodes, $\{A_1, A_2, ..., A_x\}$, at various locations within the simulated area. These adversaries employ a dynamic approach, attempting to disrupt the communication link between authenticated UAVs and the base stations by strategically positioning themselves to maximize interference.

The total jamming power affecting resource block  $k$ is defined as the cumulative contribution from all adversarial nodes as shown in eq \ref{eq:jampowersum}:

\begin{equation}
P_{\text{jam},k} = \sum_{j=1}^{x} P_{A_j,k}
\label{eq:jampowersum}
\end{equation}

where $P_{A_j,k}$ represents the jamming power from adversarial node $A_j$ at resource $k$. This jamming power depends on several factors:

\begin{equation}
P_{A_j,k} = P_{t,A_j,k} \cdot G_{A_j} \cdot L_{A_j,k}
\label{eq:jampower}
\end{equation}

where:
\begin{itemize}
    \item $P_{t,A_j,k}$ is the transmission power of adversarial node $A_j$ at subcarrier (resource block) $k$ (linear scale),
    \item $G_{A_j}$ is the antenna gain of adversarial node $A_j$ (linear scale),
    \item $L_{A_j,r,k}$ is the large-scale channel gain (accounting for path loss and shadowing) between adversarial node $A_j$ and the receiver at subcarrier $k$.
\end{itemize}

The SINR at subcarrier $k$ is then given by:

\begin{equation}
\text{SINR}_k = \frac{P_{\text{uav},k}}{P_{\text{jam},k} + \sum_{i=1}^{N_I} P_{I,i,k} + P_{N,k}} = \frac{P_{\text{uav},k}}{\sum_{j=1}^{x} P_{A_j,k} + \sum_{i=1}^{N_I} P_{I,i,k} + P_{N,k}}
\label{eq:sinr_k}
\end{equation}

where $P_{\text{uav},k}$ is the received power from the legitimate UAV, $\sum_{i=1}^{N_I} P_{I,i,k}$ represents other interference sources, and $P_{N,k}$ is the noise power, all at subcarrier $k$.

In this extended model, $P_{\text{uav},k}$ denotes the received power at resource block $k$, and $P_{I,i,k}$ and $P_{N,k}$ represent the interference and noise power per resource block, respectively. Throughout our analysis, the subscript $k$ consistently refers to the resource block index.

The thermal noise power $P_{N,k}$ is modeled as: $P_{N,k} = k_B \cdot T \cdot B \cdot NF$, where $k_B$ is Boltzmann's constant, $T$ is the system temperature, $B$ is the channel bandwidth, and $NF$ is the receiver's noise figure.

A key aspect of this extended model is the characterization of interference from multiple base stations. The interference power from each base station $P_{I,i}$ is modeled similarly to the desired signal:

\begin{equation}
P_{I,i} = P_{tx,i} + G_{total,i} - PL^{\alpha}_i(d_i, f_c) - S_i - X_{\sigma,i}
\label{eq:interference}
\end{equation}

where, $P_{tx,i}$ is the transmit power of the i-th interfering base station, $G_{total,i}$ is the combined antenna gain between the i-th interfering base station and the UAV, $PL^{\alpha}_i(d_i, f_c)$ is the path loss for the signal from the i-th interfering base station, $S_i$ accounts for small-scale fading effects, and $X_{\sigma,i}$ models shadowing specific to the channel between the i-th interfering base station and the UAV.

Packet loss calculations are done through the application of Effective SINR Mapping (ESM) techniques such as the Mutual Information Effective SINR Mapping (MIESM), that compresses the multidimensional vector of SINR values, spanning multiple resource blocks, into a single, representative effective SINR value. MIESM, in particular, leverages the concept of mutual information to achieve this compression, offering a nuanced approach to SINR aggregation. The power of these methods lies in their ability to establish a robust correlation between the computed effective SINR and the Block Error Rate (BLER), a relationship calibrated through extensive link-level simulations.\cite{nr_module}, \cite{Nr2024}. 

Received Signal Strength Indicator (RSSI):
RSSI quantifies the total received power across the allocated frequency band, including contributions from the desired signal, interference from neighboring cells, jammers, and thermal noise:

\begin{equation}
    \text{RSSI} = \sum_{k=1}^{N_{RB}} \left( P_{\text{uav},k} + P_{\text{jam},k} + \sum_{i=1}^{N_I} P_{I,i,k} + P_{N,k} \right)
\end{equation}

where ${N_{RB}}$ the number of resource blocks allocated for that UAV.
Reference Signal Received Power (RSRP): 
RSRP is defined as the average power received from the reference signals transmitted by the serving base station. Unlike RSSI, RSRP includes only the power of specific reference signal elements, excluding interference and noise:

\begin{equation}
    \text{RSRP} = \frac{1}{N_{RS}} \sum_{k=1}^{N_{RS}} P^{RS}_{rx,k}
\end{equation}

where $N_{RS}$ is the number of resource elements carrying reference signals, and $P^{RS}_{rx,k}$ denotes the received power of the $k$-th reference signal element.

\section{Anti-Jamming Recovery Framework}
\label{sec:Anti_Jamming}
To combat the jamming effects described in the previous sections, we propose a comprehensive recovery framework that leverages four key mechanisms: adaptive beamforming, selective notching, bandwidth part selection, and intelligent retransmission. The recovery problem can be formulated as:
\begin{equation}
\mathcal{R} = \{\mathbf{w}^*, \mathbf{n}_{\text{notch}}, B_{\text{idx}}, \mathbf{r}\}
\end{equation}

where $ \mathbf{w}$ represents the beamforming configuration, $\mathbf{n}$ denotes the notching pattern, $B_{\text{idx}}$ indicates bandwidth part selection, and $\mathbf{r}$ encapsulates the retransmission strategy.

\subsection{Adaptive Beamforming}

The beamforming vector $ \mathbf{w} \in \mathbb{C}^M $ is designed to maximize the SINR in the presence of jamming:
\begin{equation}
\mathbf{w}^* = \arg\max_{\mathbf{w}} \frac{|\mathbf{h}^H\mathbf{w}|^2}{\sum_{i=1}^{N_I} |\mathbf{g}_i^H\mathbf{w}|^2 + |\mathbf{j}^H\mathbf{w}|^2 + \sigma^2\|\mathbf{w}\|^2}
\end{equation}
where $\mathbf{h}$ represents the channel vector to the legitimate receiver, $\mathbf{g}_i$ is the channel vector to the $i$-th interferer, $\mathbf{j}$ is the channel vector to the jammer, and $\sigma^2$ is the noise power.

In this context, $M$ refers to the number of antenna elements in the UAV or gNB antenna array. Specifically, the beamforming vector $ \mathbf{w} $ consists of $M$ complex weights, each corresponding to one of the $M$ antenna elements, which control the phase and amplitude of the transmitted or received signals. This allows for spatial filtering to mitigate interference and jamming effects.

In practice, we parameterize the beamforming strategy through azimuth and zenith angles:
\begin{equation}
\mathbf{a}_{\text{bf}} = [\theta_{\text{UAV}}, \phi_{\text{UAV}}, \theta_{\text{gNB}}, \phi_{\text{gNB}}]
\end{equation}

The relationship between these angles and the beamforming vector is given by:
\begin{equation}
\mathbf{w}^*(\mathbf{a}_{\text{bf}}) = \frac{1}{\sqrt{M}}[e^{j\mathbf{k}(\mathbf{a}_{\text{bf}})^T\mathbf{p}_1}, e^{j\mathbf{k}(\mathbf{a}_{\text{bf}})^T\mathbf{p}_2}, \ldots, e^{j\mathbf{k}(\mathbf{a}_{\text{bf}})^T\mathbf{p}_M}]^T
\end{equation}

where $\mathbf{k}(\mathbf{a}_{\text{bf}}) = \frac{2\pi}{\lambda}[k_x, k_y, k_z]^T$ with $k_x = \sin\theta_{\text{UAV}}\cos\phi_{\text{UAV}}$, $k_y = \sin\theta_{\text{UAV}}\sin\phi_{\text{UAV}}$, and $k_z = \cos\theta_{\text{UAV}}$ is the wave vector, and $\mathbf{p}_m$ is the position of the $m$-th antenna element in the array. While beamforming is not directly embedded in the SINR equation, it implicitly improves $P_{\text{uav},k}$ and may reduce $P_{\text{jam},k}$ and $P_{I,i,k}$ through directional gain control.

\subsection{Selective Spectrum Notching}
The notching vector $\mathbf{n} = [n_1,\ldots,n_{N_{RB}}]^T \in [0,1]^{N_{RB}}$ indicates the degree to which each resource block is notched. The effective SINR at resource block  $k$ after notching is:
\begin{equation}
\text{SINR}_k = \frac{(1-n_k)P_{\text{uav},k}}{(1-\eta_{\text{notch}} n_k)P_{\text{jam},k} + \sum_{i=1}^{N_I} P_{I,i,k} + P_{N,k}}
\end{equation}
where $\eta_{\text{notch}} \in [0,1]$ represents the notching efficiency factor.
The notching strategy is parameterized as:
\begin{equation}
\mathbf{n}_{\text{notch}} = [RB_{\text{start}}, RB_{\text{num}}, I_{\text{notch}}]
\end{equation}
This is mapped to the notching vector as follows:
\begin{equation}
n_k = 
\begin{cases} 
I_{\text{notch}} & \text{if } k \in [RB_{\text{start}}, RB_{\text{start}} + RB_{\text{num}} - 1] \\
0 & \text{otherwise}
\end{cases}
\end{equation}

\subsection{Bandwidth Part Selection}
The concept involves dynamically adjusting the bandwidth either increasing or decreasing it to improve resilience against interference and jamming. This adaptability enhances robustness in multiple ways:

\begin{itemize}
    \item \textbf{Bandwidth reduction:} When jamming is localized in specific frequency regions, reducing the transmission bandwidth to avoid those affected areas can improve overall signal quality, even at the cost of reduced peak throughput.
    
    \item \textbf{Bandwidth increase:} In scenarios where narrowband jamming is present, spreading the communication over a wider bandwidth can exploit frequency diversity. This effectively ``dilutes'' the impact of the jammer across more subcarriers, reducing its effectiveness.
    
    \item \textbf{Strategic bandwidth selection:} The framework supports the selection of specific bandwidth parts (BWPs), each defined by a central frequency and bandwidth. BWPs can be chosen to maximize spectral distance from known jamming frequencies, thereby enhancing link reliability.
\end{itemize}

Formally, we denote the index of the selected bandwidth part as $B_{\text{idx}} \in \{1, 2, \ldots, B_{\text{max}}\}$, where each BWP $b$ is characterized by its center frequency $f_c^b$ and bandwidth $W_b$. We modify these BWP parameters over time using discrete indexes to dynamically adapt to the changing jamming environment.

\subsection{Retransmission Strategy}

The retransmission strategy $\mathbf{r} = [r_{\text{max}}]$ consists of the maximum number of retransmissions $r_{\text{max}} \in \{0, 1, 2, \ldots, R_{\text{max}}\}$. The packet error rate after $r$ retransmissions is:

\begin{equation}
\text{PER}(r) = 1 - (1-\text{BLER})^{r+1}
\end{equation}

where BLER is the Block Error Rate for a single transmission.

\subsection{Problem Formulation}
\label{sec:problem_formulation}
We formulate the anti-jamming communication problem between an authenticated UAV and a base station as a constrained optimization task. The objective is to minimize communication inefficiencies under jamming, including packet loss, retransmission attempts, latency, and jitter. These are mitigated through joint optimization of beamforming, spectrum notching, and bandwidth allocation.
The problem is defined as:
\begin{equation}
\begin{aligned}
\mathbf{P1}: \quad \min_{\mathbf{w}, \mathbf{n}, B_{\text{idx}}} \quad & 
\Phi P_L + \beta R_a + \mu L + \Psi J \\
\text{s.t.} \quad 
& B_{\text{idx}} \in \{1, 2, \ldots, B_{\text{max}}\}, \\
& \mathbf{C}_{\text{min}} \preceq \mathbf{C}(\mathbf{w}, \mathbf{n}, B_{\text{idx}}) \preceq \mathbf{C}_{\text{max}}, \\
& \|\mathbf{w}\|^2 \leq 1, \\
& 0 \leq n_k \leq 1, \quad \forall k \in \{1,\ldots,N_{RB}\}, \\
& \sum_{k=1}^{N_{RB}} n_k \leq N_{\text{notch}}^{\text{max}}, \\
& L \leq L_{\text{max}}, \quad J \leq J_{\text{max}}.
\end{aligned}
\end{equation}
Here:

\begin{itemize}
\item $\mathbf{w} \in \mathbb{C}^M$ is the beamforming vector,
\item $\mathbf{n} \in [0,1]^{N_{RB}}$ is the notching vector across $N_{RB}$ resource blocks,
\item $B_{\text{idx}} \in \{1, 2, \ldots, B_{\text{max}}\}$ is the index of the selected bandwidth part,
\item $N_{\text{notch}}^{\text{max}}$ limits the total number of notched resource blocks.

$\mathbf{C}(\cdot): \mathbb{C}^M \times \mathbb{R}^{N_{RB}} \times \mathbb{Z}_+ \rightarrow \mathbb{R}^D$ captures system-level metrics such as SINR, interference levels, and coverage quality, where $D$ is the dimension of the constraint vector.

The performance metrics include:
\item  Packet loss rate $P_L \in [0,1]$, calculated from the BLER after applying ESM techniques,
\item  Retransmission attempts $R_a \in \mathbb{N}_0$, derived from the retransmission strategy $\mathbf{r}$,
\item  End-to-end latency $L \in \mathbb{R}_+$, which accounts for processing and transmission delays,
\item Jitter $J \in \mathbb{R}_+$, representing the variation in packet arrival times.

\end{itemize}
\vspace{1em}
\section{Cross-Layer Configuration for Jamming Resilience}

This work proposes a reinforcement learning framework for optimizing 5G cross-layer configurations against jamming threats.

\begin{figure*}[htbp]
    \centering
    \captionsetup{font=footnotesize}
    \includegraphics[width=0.98\textwidth]{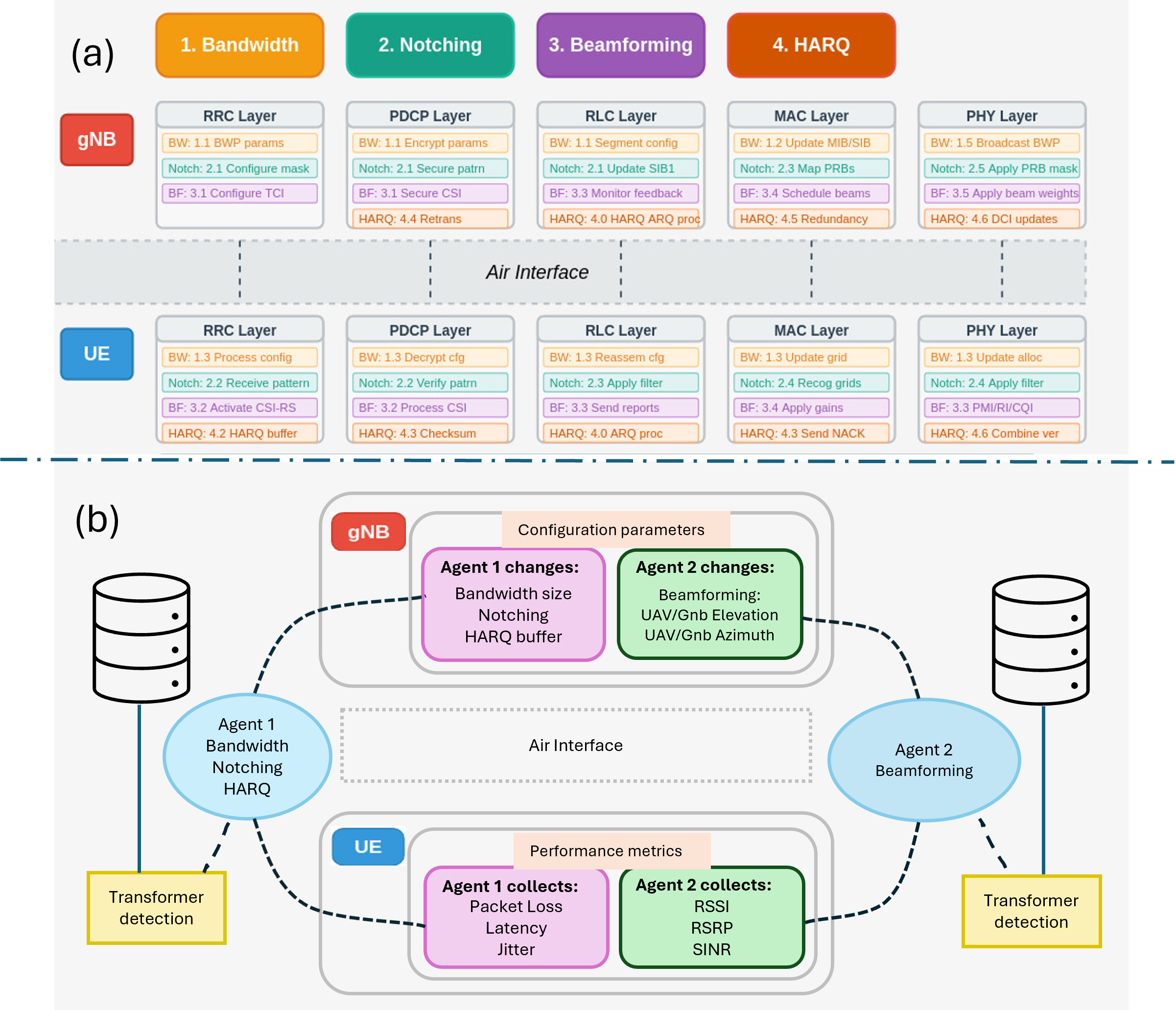} 
    \caption{5G Network Parameter Configuration Procedures. (a) Conventional procedure showing the sequential configuration of Bandwidth (yellow), Notching (green), Beamforming (purple), and HARQ ( orange) parameters across protocol layers for both gNB and UE components. (b) MAPPO-based procedure using dual agents: Agent 1 (pink) executes bandwidth, notching, and HARQ configuration procedures while Agent 2 (green) implements beamforming adjustments, with both continuously collecting performance metrics to optimize configurations.}
    \label{fig:2}
\end{figure*}

Figure \ref{fig:2} (a) illustrates how standard 5G implementations coordinate configurations across multiple protocol layers. The process follows four key mechanisms:

\begin{enumerate}
   \item \textbf{Bandwidth adjustments} (orange blocks): Starting with BWP setup at the RRC layer, this mechanism flows through the protocol stack to configure grid updates and Physical Resource Block (PRB) allocation at both gNB and UE sides.
   
   \item \textbf{Spectrum notching} (green blocks): This technique implements frequency avoidance through mask configuration, secure path provisioning, PRB mapping, and filtering operations across the protocol layers.
   
   \item \textbf{Beamforming} (purple blocks): On the gNB side, this spatial processing begins with Transmission Configuration Indicator (TCI) configuration and proceeds through scheduling and weight updates. Meanwhile, the UE handles (Channel State Information Reference Signal) CSI-RS activation, processing, and gain control to complete the spatial adaptation loop.
   
   \item \textbf{HARQ mechanisms} (brown blocks): This error recovery process manages retransmissions through Downlink Control Information (DCI) updates, redundancy versioning, and ARQ buffering, ensuring reliable communications despite potential packet losses.
\end{enumerate}

In Figure \ref{fig:2} (b), these procedures are optimized using a MAPPO framework. The first agent manages bandwidth selection, notching, and HARQ buffer operations, guided by real-time network metrics (packet loss, latency, jitter). The second agent tunes beamforming parameters (azimuth/elevation) based on signal quality metrics such as RSSI, RSRP, and SINR. Together, the agents autonomously coordinate cross-layer adaptation to maintain robust performance under jamming conditions.

\section{Multi-Agent Approach for Anti-Jamming Recovery}

\label{sec:mappo_framework}
MAPPO extends the Proximal policy optimization (PPO) algorithm to multi-agent scenarios, making it well-suited for our anti-jamming recovery problem. Each agent $i \in \{1,2\}$ learns a policy $\pi_{\theta_i}$ parameterized by $\theta_i$ to maximize its expected cumulative return:

\begin{equation}
V_i(\theta_i) = \mathbb{E}_{\tau \sim \pi_{\theta_i}}\left[\sum_{t=0}^{T} \gamma^t R_i(s_t, \mathbf{a}_t)\right]
\end{equation}

Where $\tau = (s_0, \mathbf{a}_0, R_0, \ldots, s_T)$ represents a trajectory of states, joint actions, and rewards, $R_i(s_t, \mathbf{a}_t)$ is the reward obtained by agent $i$ at time step $t$, $\gamma \in [0, 1)$ is the discount factor, and $T$ is the time horizon.

To enhance coordination between agents, MAPPO incorporates a proximity term that penalizes large policy deviations between neighboring agents:

\begin{equation}
\tilde{V}_i(\theta_i) = V_i(\theta_i) - \chi \sum_{j \neq i} D_{\text{KL}}(\pi_{\theta_i}(\cdot | s_{i,t}) \| \pi_{\theta_j}(\cdot | s_{j,t}))
\end{equation}

Where $\chi > 0$ is a hyperparameter controlling the proximity penalty strength, and $D_{\text{KL}}$ represents the Kullback-Leibler divergence measuring the difference between policies.

The policy update mechanism follows the PPO clipped objective function:

\begin{equation}
\Xi^{\text{CLIP}}_i(\theta_i) = \hat{\mathbb{E}}_t \left[ \min\left(\rho_t(\theta_i)\hat{adv}_i^t, \text{clip}\left(\rho_t(\theta_i), 1-\epsilon, 1+\epsilon\right)\hat{adv}_i^t\right) \right]
\end{equation}

Where $\rho_t(\theta_i) = \frac{\pi_{\theta_i}(\mathbf{a}_{i,t}|s_{i,t})}{\pi_{\theta_i^{\text{old}}}(\mathbf{a}_{i,t}|s_{i,t})}$ is the probability ratio, $\hat{A}_i^t$ is the advantage estimate, and $\epsilon$ is the clipping parameter.

\subsection{FAIR WINGS System Architecture}
\label{sec:FAIR}

In our FAIR WINGS (Framework for AI-Resilient Wireless Networks in Ground UAV Systems), we implement a UAV anti-jamming system that optimizes reception capabilities in 5G networks by considering multiple factors simultaneously: beam configuration, transmission parameters, jamming characteristics, network connectivity, and mobility constraints. The system leverages key 5G performance metrics (SINR, RSSI, RSRP, packet loss rate, latency, and jitter) as inputs and employs two specialized agents in a multi-agent reinforcement learning framework:

\begin{enumerate}
    \item Agent 1 (Bandwidth Agent): Manages bandwidth part selection, notching patterns, and HARQ buffer settings
    \item Agent 2 (Beamforming Agent): Controls beamforming angles and elevation for both UAV and gNB
\end{enumerate}

This division of responsibilities enables specialized learning and coordinated action across different dimensions of the anti-jamming problem. The agents are a third entity that can run in either of the devices, the UE or the gNB, but should have access to the other device's configuration. In our simulation, this is accomplished by exchanging the parameters between both devices as illustrated in figure \ref{fig:2}.

Additionally, our final system incorporates a transformer-based jamming detector operating in inference mode. This transformer provides classification logits for both benign network conditions and attack scenarios directly to the MAPPO agents' observation space.

\subsubsection{State Space Formulation}
The state space $\mathcal{S}$ incorporates the UAV's communication parameters and environmental metrics:
\begin{equation}
s_t = \{s_{1,t}, s_{2,t}\} \in \mathcal{S}
\end{equation}
Where:
\begin{itemize}
    \item $s_{1,t} = s_t^{\text{perf}}$ captures network performance metrics for Agent 1: packet loss, latency, and jitter measurements
    \item $s_{2,t} = s_t^{\text{sig}}$ captures signal quality metrics for Agent 2: SINR, RSSI, and RSRP measurements
\end{itemize}

In our implementation, we distribute these observations across the two agents based on their responsibilities. Agent 1 receives the network performance metrics ($s_{1,t}$), which directly impact its decisions regarding bandwidth allocation, notching, and HARQ buffer management. Meanwhile, Agent 2 receives signal quality metrics ($s_{2,t}$), which inform its beamforming decisions. This division allows each agent to focus on the most relevant information for its specific task.

\subsubsection{Hybrid Action Space}
The action space employs a hybrid representation distributed across the two agents:

\paragraph{Agent 1: Bandwidth, Notching, and HARQ Management}
\begin{equation}
\mathcal{A}_1 = \{\mathbf{a}_{1,t} = [\mathbf{a}_{\text{notch}}, B_{\text{idx}}, r_{\text{max}}] \} \subset \mathbb{Z}^3 \times \mathbb{Z} \times \mathbb{Z}^+
\end{equation}

Where:
\begin{itemize}
\item $\mathbf{a}_{\text{notch}} = [RB_{\text{start}}, RB_{\text{num}}, I_{\text{notch}}]$ represents the notching strategy parameters
\item $B_{\text{idx}} \in \mathbb{Z}$ is the bandwidth part index selection
\item $r_{\text{max}} \in \mathbb{Z}^+$ is the maximum number of retransmission attempts (HARQ buffer size)
\end{itemize}

Our implementation executes these spectrum management actions through two key mechanisms. First, resource block notching applies masks to the MAC scheduler that prevent transmission on specific resource blocks experiencing jamming. The agent determines which portions of the spectrum to avoid by selecting a starting resource block index and the number of consecutive blocks to notch. The notching indicator allows the agent to toggle this feature on or off based on jamming conditions.

Second, bandwidth part switching enables adaptation between different numerologies. The system supports two bandwidth parts: BWP 0 with numerology 2 (using 100 MHz bandwidth) and BWP 1 with numerology 3 (using 50 MHz bandwidth). Higher numerology provides lower latency but less coverage, allowing the agent to make trade-offs based on current conditions. The BWP index selection action determines which configuration to use.

\paragraph{Agent 2: Beamforming Control}
\begin{equation}
\mathcal{A}_2 = \{\mathbf{a}_{2,t} = \mathbf{a}_{\text{bf}} \} \subset \mathbb{R}^4
\end{equation}

Where:
\begin{itemize}
\item $\mathbf{a}_{\text{bf}} = [\theta_{\text{UAV}}, \phi_{\text{UAV}}, \theta_{\text{gNB}}, \phi_{\text{gNB}}]$ represents the azimuth and zenith angles for UAV and gNB antennas
\end{itemize}

Our implementation uses an ideal beamforming helper with a custom beamforming algorithm that accepts these angle parameters. The beamforming actions configure both the UAV and gNB antenna orientations to establish the optimal communication link that minimizes interference. The agent can set azimuth angles (horizontal plane orientation) and zenith angles (vertical plane orientation) for both endpoints of the communication link.

The complete action space is thus $\mathcal{A} = \mathcal{A}_1 \times \mathcal{A}_2$, with each agent responsible for its respective subspace.

\subsubsection{Reward Function Design}
The reward function $R_i: \mathcal{S} \times \mathcal{A} \rightarrow \mathbb{R}$ for each agent $i$ balances multiple performance objectives and is formulated as:
\begin{equation}
R_i(s_t, \mathbf{a}_t) = 2 \cdot \left(\sum_{j} w_{i,j} \cdot m_j\right) - 1
\end{equation}

Where $m_j$ represents normalized performance metrics and $w_{i,j}$ their respective weights for agent $i$. Our implementation uses agent-specific reward functions to reflect the different optimization objectives of each agent in our multi-agent system:

\begin{table}[h]
\centering
\begin{tabular}{lcc}
\hline
\textbf{Metric} & \textbf{Agent 1} & \textbf{Agent 2} \\
\hline
Packet Loss ($1-p_t$) & 0.4 & 0.4 \\
SINR ($S_t$) & 0.2 & 0.4 \\
RSRP ($R_{s,t}$) & 0.0 & 0.2 \\
Latency ($L_t$) & 0.2 & 0.0 \\
Jitter ($J_t$) & 0.2 & 0.0 \\
\hline
\end{tabular}
\caption{Agent-specific reward weights for different performance metrics}
\label{tab:reward_weights}
\end{table}

The metrics are defined as follows:
\begin{itemize}
\item $p_t \in [0, 1]$ is the packet loss rate, with $1-p_t$ representing successful packet delivery
\item $S_t \in [0, 1]$ is the normalized SINR
\item $R_{s,t} \in [0, 1]$ is the normalized RSRP
\item $L_t \in [0, 1]$ is the normalized latency (inverted so lower is better)
\item $J_t \in [0, 1]$ is the normalized jitter (inverted so lower is better)
\end{itemize}

Our implementation uses a dynamic normalization approach that tracks minimum and maximum values for each metric to ensure they have comparable scales:
\begin{equation}
\text{Normalize}(k, var) = \frac{<var> - \min_k}{\max_k - \min_k}
\end{equation}

This normalization maps all performance indicators to the $[0,1]$ range, with higher values representing better performance. For latency and jitter, the normalized values are inverted (i.e., $1.0 - \text{Normalize}(k, v)$) since lower raw values are preferable. The weighted sum is then scaled to the $[-1,1]$ range and clipped to prevent extreme values, providing appropriate gradients for the reinforcement learning algorithm.

The design of our reward function reflects the different responsibilities of each agent. Agent 2 (Beamforming Agent) focuses on optimizing signal quality metrics (SINR and RSRP) alongside packet loss, while Agent 1 (Bandwidth Agent) balances packet loss with latency and jitter considerations. This specialization enables more effective learning and coordination in our multi-agent system, as each agent can focus on the aspects of performance most relevant to its control parameters.

\subsubsection{Learning Process and Integration}

The agents learn through repeated interactions with the environment. At each timestep $t$:
\begin{enumerate}
\item Each agent $i \in \{1,2\}$ receives its state information $s_{i,t}$
\item Agents select actions according to their policies: $\mathbf{a}_{i,t} \sim \pi_{\theta_i}(\cdot|s_{i,t})$
\item The environment transitions to state $s_{t+1}$ according to $\mathcal{P}(s_{t+1}|s_t,\mathbf{a}_t)$ where $\mathbf{a}_t = [\mathbf{a}_{1,t}, \mathbf{a}_{2,t}]$
\item Agents receive rewards $R_i(s_t, \mathbf{a}_t)$ based on the new state
\item Policy parameters are updated following the MAPPO algorithm
\end{enumerate}
We noticed that even without jamming in the network can be further optimized. 

\subsection{Detection-Enhanced Transformer FAIR-WINGS (DET-FAIR-WINGS)}
\label{sec:transformer}

Following our experiment, we integrated the transformer model available in \cite{viana2025pcafeatured} to our recovery system and renamed it to DET-FAIR-WINGS framework. The transformer combines a U-shaped transformer architecture with the Principal component analysis (PCA) feature enhancement for efficient jamming detection in 5G UAV networks.

Our model uses a U-shaped structure with:
\begin{itemize}
\item Three encoder stages (256→128→64)
\item Three decoder stages (64→128→256)
\item Skip connections between matching layers
\end{itemize}
Each processing block combines normalization, convolution, multi-head attention, and activation functions.
\subsubsection{Dataset}
We used the synthetic jamming dataset from \cite{SyntheticJViana}, which includes various wireless attack scenarios in different network conditions.
\subsubsection{Feature Processing}
Our method processes RSSI and SINR signals by:
\begin{enumerate}
\item Using 300-sample windows
\item Applying nine different transformations
\item Using PCA to retain key components
\item Selecting the five most important features
\item Combining with original signals
\end{enumerate}
This creates 90 features for LoS and 54 for NLoS scenarios. More information available in \cite{viana2025pcafeatured}.
\subsubsection{Model Training}
We trained our detection model using an uncertainty-aware loss function:
\begin{equation}
Loss = \frac{Loss_{primary} - \alpha_{uncertainty}E(P)}{G_{steps}}
\end{equation}
where E(P) measures prediction uncertainty.

\subsubsection{MAPPO Integration}
For our MAPPO reinforcement learning agents, we utilize the pre-trained transformer in inference mode. The MAPPO agents receive two critical classification logits from the transformer:

\begin{itemize}
\item \textbf{Class 1 Logits ($l_1$):} The raw classification scores from the transformer for benign network conditions, providing confidence levels for normal operation detection.
\item \textbf{Class 2 Logits ($l_2$):} The raw classification scores for attack scenarios, offering granular information about potential jamming activities.
\end{itemize}

These pre-softmax logits provide the MAPPO framework with direct uncertainty information about potential jamming activities, enabling more robust decision-making without requiring further training of the transformer itself. The agents are trained to optimize network performance by incorporating these detection model outputs in their observation space, enhancing resilience against jamming.

\section{Performance Results}
\label{sec:results}
In this section, we present the performance results of our resilience evaluation across four scenarios: general transmission, single agent, multiagent, and multiagent transformer detection. Unless explicitly mentioned, the network configuration and their corresponding values are described in Table~\ref{tab2}, and the learning parameters are provided in Table~\ref{tab3}.

\begin{table}[ht]
\centering
\begin{threeparttable}
\begin{tabular}{@{}ll@{}}
\toprule
Scenario Parameters & Value  \\ \midrule
Terrestrial Users   & $0,3,5,10$  \\
Authenticated UAVs  & 1  \\
Small Cells  & 2   \\c
Small cell height &  10 m \\
Attackers   & $0,1,2$ \\
Speeds  & 10 m/s    \\
Modulation scheme        & OFDM \\
Small cell power         & 4 dBm  \\
Authenticated UAV power  & 2 dBm \\
Attackers power    & 0,2,5,10 dBm \\
Authenticated UAV  position &  URD*  \\
Attackers position & URD* \\
Small cells position & URD* \\
Scenario  &  UMi  \\
Distance & $100, 200$ m \\
Simulation time & 10 s \\
\bottomrule
\end{tabular}
\begin{tablenotes}
  \small
  \item   *URD - Uniformly Random Distributed
\end{tablenotes}
\end{threeparttable}
\caption{Network Parameters.}
\label{tab2}
\end{table}

\begin{table}[!ht]
\centering
\begin{tabular}{@{}lll@{}}
\toprule
%\cmidrule{1-2} 
 Learning Configuration & Value \\ \midrule
 Network structure for actor & [input\_dims, 128, 128, n\_actions] \\
 Network structure for critic & [input\_dims, 128, 128, 1] \\
 Number of training episodes & 49 \\
 Learning rate for actor/critic & 1e-3 to 1e-5 (scheduled) \\
 Reward discount & 0.95 \\
 Batch size &  32 to 1024 (scheduled) \\
 Number of epochs & 200000 \\
\bottomrule
\end{tabular}
\caption{Network Parameters.}
\label{tab3}
\end{table}

\subsection{Convergence Analysis}

\begin{figure}[htbp]
    \centering
    \captionsetup{font=footnotesize}
    \includegraphics[width=0.48\textwidth]{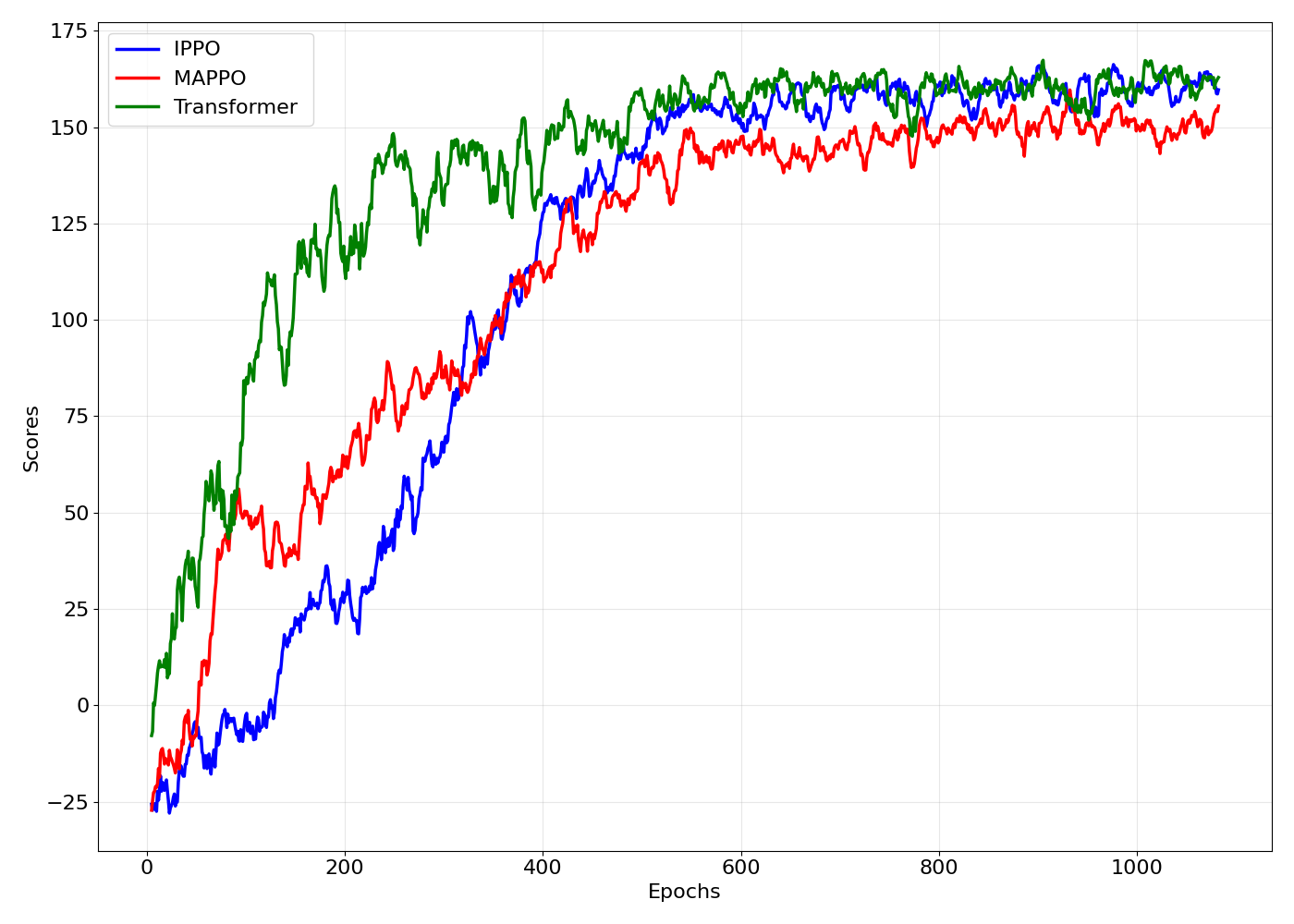} 
    \caption{Learning Rate: Training progression of PPO, MAPPO, and MAPPO+transformer approaches in terms of cumulative reward over epochs. The transformer-based approach (green line) demonstrates significantly faster convergence compared to the alternatives.}
    \label{fig:3}
\end{figure}

Figure~\ref{fig:3} presents the convergence charts comparing PPO, MAPPO, and MAPPO+transformer approaches. The plots illustrate the training progression of each algorithm in terms of cumulative reward over training epochs. As evident from the charts, the MAPPO+transformer (labeled as ``Transformer'' in the figure) demonstrates significantly faster convergence compared to the baseline PPO and standard MAPPO implementations.
Specifically, the transformer-based approach reaches a score of approximately 125 in just 200 epochs, while MAPPO and PPO remain below 50 at this point. Most notably, MAPPO+transformer achieves stable high performance in only about 400 epochs, while PPO and MAPPO require nearly twice as many epochs to reach comparable performance levels.
The superior convergence rate of the MAPPO+transformer can be attributed to its enhanced ability to capture long-range dependencies in the state space through the self-attention mechanism. 
Despite theoretical expectations that incorporating transformer architectures would increase training time, our experiments revealed a surprising outcome. Although adding transformer modules to MAPPO introduces more parameters and computational complexity per iteration, we observed that the overall training time decreased. This counterintuitive result can be attributed to the transformer's superior ability to efficiently process and extract relevant patterns from the observation space.

\subsection{Packet Loss Comparison}

\begin{figure}[htbp]
    \centering
    \captionsetup{font=footnotesize}
    \includegraphics[width=0.48\textwidth]{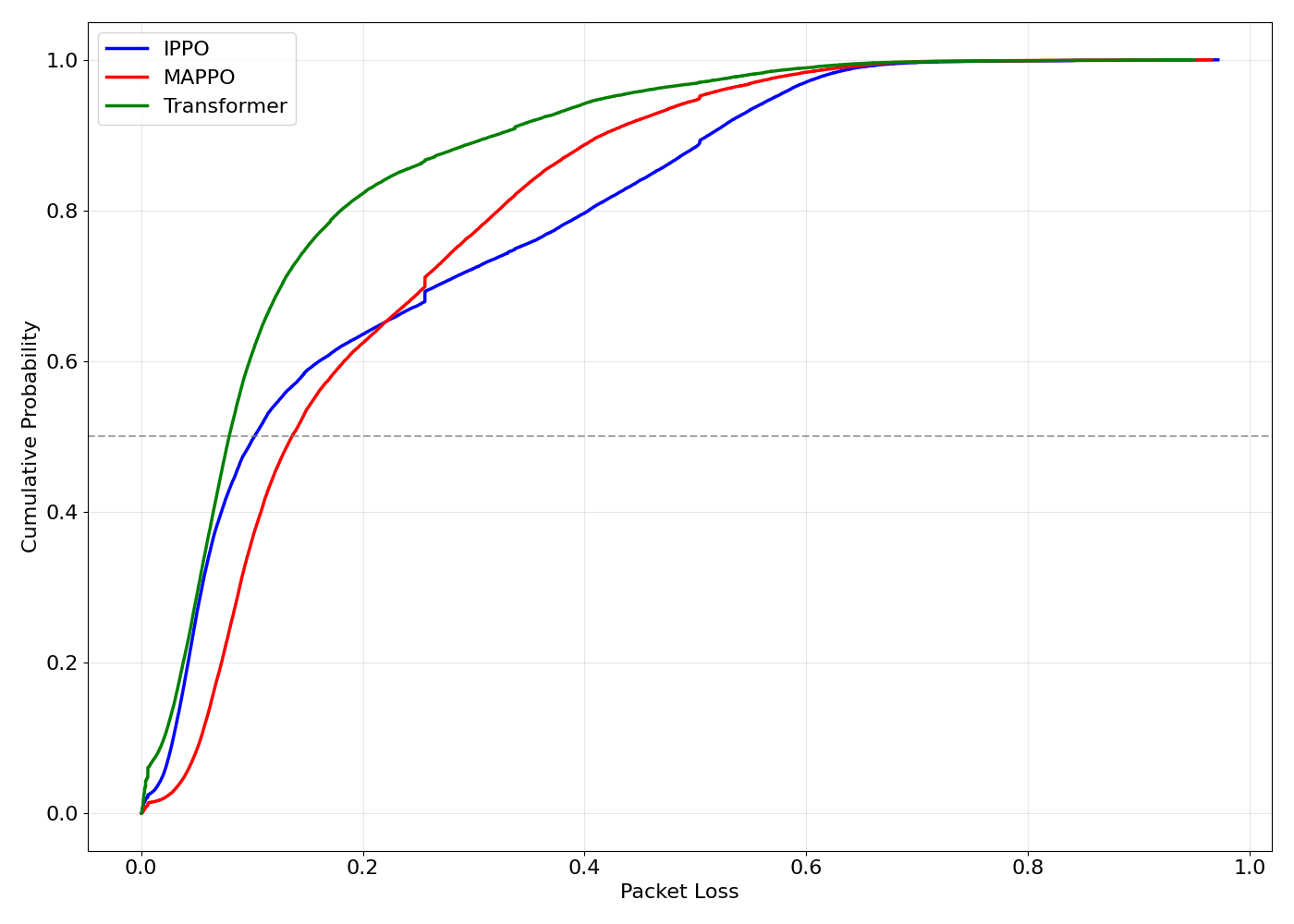} 
    \caption{Packet loss moving average comparison between IPPO, MAPPO, and MAPPO+transformer approaches over training steps. The straight lines represent the overall tendencies of the packet loss rates.}
    \label{fig:4}
\end{figure}

Figure~\ref{fig:4} presents a comparative analysis of packet loss rates through a cumulative distribution function (CDF) across the three different approaches. The CDF provides insights into the probability distribution of packet loss rates encountered during operation. The MAPPO+transformer approach (green line) shows superior performance, with approximately 80\% of its packet loss measurements falling below 0.2, compared to roughly 65\% for PPO and MAPPO at the same threshold. This indicates that the transformer-based approach maintains lower packet loss rates more consistently.
The results clearly demonstrate the progressive improvement in network resilience when using the transformer-based approach. While the baseline PPO approach (blue line) shows competitive performance in some regions of the distribution, the MAPPO+transformer consistently outperforms both alternatives across the entire packet loss range, particularly in the critical mid-range (0.2-0.4) where network performance significantly impacts user experience.

\section{Conclusion}
\label{sec:conclusion}

We presented DET-FAIR-WINGS, a framework that combines transformer-based jamming detection with multi-agent reinforcement learning to protect 5G UAV networks against jamming attacks. Our approach integrates transformer classification logits directly into the MAPPO agents' observation space, enabling more effective and responsive defense strategies.
Our evaluation demonstrated three key findings. First, incorporating transformer models in inference mode with their classification logits ($l_1$ and $l_2$) as part of the MAPPO agents' observation space enhanced jamming detection capabilities, allowing for proactive rather than reactive defense mechanisms. 
Second, our MAPPO+transformer approach exhibited sample efficiency, converging substantially faster than both standard MAPPO and PPO implementations. As demonstrated in our convergence analysis, the transformer-enhanced approach reached comparable performance levels in approximately one-third of the training time required by conventional methods. This acceleration was consistently observed across diverse network configurations and jamming scenarios, highlighting the approach's robustness.
Third, we empirically validated the practical benefits of our framework through comprehensive packet loss analysis. The CDF analysis revealed that the MAPPO+transformer approach maintained significantly lower packet loss rates across the entire distribution range, with particularly notable improvements in the critical mid-range areas that most directly impact network quality of service. These improvements were corroborated by the packet loss trends over training steps, where our approach demonstrated consistently superior performance throughout the training process.
The DET-FAIR-WINGS framework advances UAV communication security by simultaneously optimizing multiple network parameters: beamforming alignment, retransmission mechanisms, bandwidth allocation, and interference mitigation. Our findings challenge previous assumptions about the feasibility of implementing sophisticated multi-agent reinforcement learning techniques in 5G communications, demonstrating that with appropriate design considerations, these approaches can be both practical and effective.
Future research directions could explore scaling this approach to larger and more heterogeneous networks and exploring the potential of transformer-based metrics as replacements for traditional interference indicators like RSSI and SINR represents a promising avenue for further investigation in next-generation wireless security frameworks and new studies on the improvements in the general network. 

\section*{Acknowledgment}
This work was partly funded by Project SOFIA-AIR (PID2023-147305OB-C31) (MICIU /10.13039/501100011033 / AEI / EFDR, UE).

% \printbibliography

\vskip -2\baselineskip plus -1fil

\begin{IEEEbiography}[{\includegraphics[width=1in, height=1.25in, clip, keepaspectratio]
{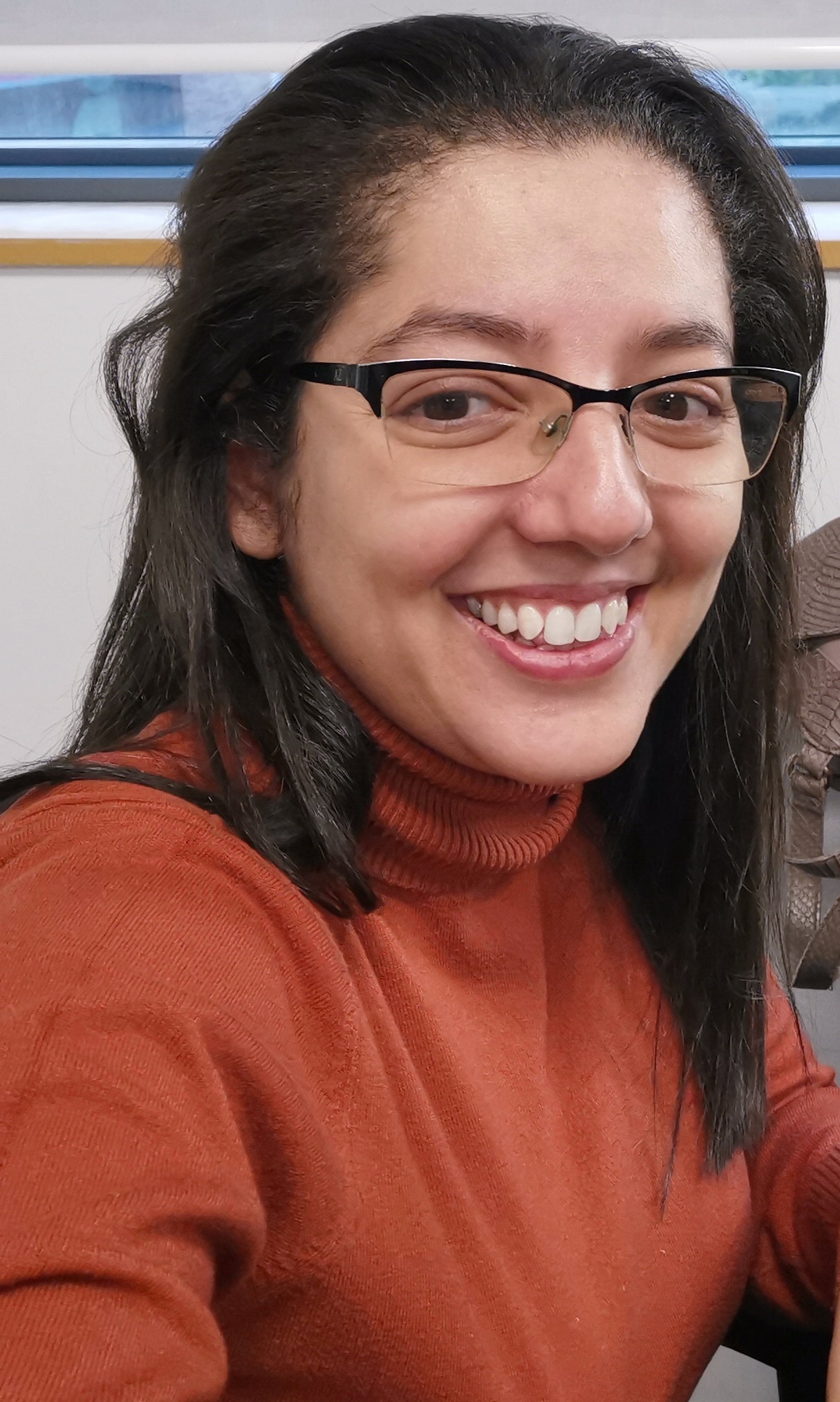}}]{Joseanne Viana} 
is a Ph.D. researcher at UC3M - Charles III University of Madrid. She received her bachelor’s degree in telecommunication engineering from the University of Campinas, Brazil. She is an Early-Stage Researcher in the project TeamUp5G, a European Training Network under the MSCA ITN of the European Commission’s Horizon 2020. Her research interests include wireless communications applied to interconnected systems such as UAVs, aerial vehicles, and non-terrestrial devices.
\end{IEEEbiography}

 \vskip -2\baselineskip plus -1fil
\begin{IEEEbiography}[{\includegraphics[width=1in, height=1.25in, clip, keepaspectratio]{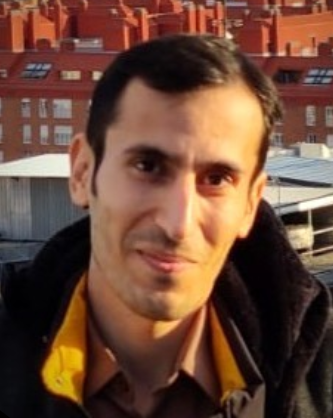}}]{Hamed Farkhari} 
is a Ph.D. researcher at ISCTE - Lisbon University Institute. He serves as an Early-Stage Researcher in the TeamUp5G group, a European Training Network under the Marie Skłodowska-Curie Actions (MSCA ITN) of the European Commission’s Horizon 2020 program. His research interests and work focus on cybersecurity, machine learning, deep learning, data science, meta-heuristic techniques, and optimization algorithms.
\end{IEEEbiography}
\vskip -2\baselineskip plus -1fil

\begin{IEEEbiography}[{\includegraphics[width=1in,height=1.25in,clip,keepaspectratio]{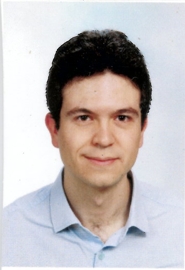}}]{Víctor P. Gil Jiménez} (Senior Member, IEEE) received a B.S. degree (Hons.) in Telecommunications from the University of Alcalá in 1998 and an M.S. degree (Hons.) in Telecommunications and a Ph.D. degree (Hons.) from Universidad Carlos III de Madrid in 2001 and 2005, respectively. He visited the University of Leeds, U.K., in 2003, Chalmers Technical University, Sweden, in 2004, and the Instituto de Telecomunicações, Portugal, from 2008 to 2010. He is an Associate Professor with the Department of Signal Theory and Communications, Universidad Carlos III de Madrid. He has led private and national Spanish projects and participated in European and international projects. He holds one patent and has published over 80 journal articles/conference papers and nine book chapters. His research interests include advanced multicarrier systems for wireless radio, satellite, and visible light communications. He was the IEEE Spanish Communications and Signal Processing Joint Chapter Chair from 2015 to 2023. Thesis Award from the Professional Association of Telecommunication Engineers of Spain in 1998 and 2006, respectively. 
\end{IEEEbiography}

\EOD

\clearpage  


\begin{thebibliography}{99}

\bibitem{ref1}
G. Geraci, A. Garc\'{\i}a-Rodr\'{\i}guez, M. M. Azari, A. Lozano, M. Mezzavilla, S. Chatzinotas, Y. Chen, S. Rangan, and M. Di Renzo,
``What Will the Future of UAV Cellular Communications Be? A Flight from 5G to 6G,''
arXiv preprint arXiv:2105.04842,
2021.

\bibitem{Skokowski2024}
P. Skokowski, K. Malon, M. Kryk, K. Ma\'{s}lanka, J. M. Kelner, P. Rajchowski, and J. Magiera,
``Practical Trial for Low-Energy Effective Jamming on Private Networks With 5G-NR and NB-IoT Radio Interfaces,''
IEEE Access, vol. 12, pp. 51523--51535,
2024.

\bibitem{ref2}
A. S. Abdalla, K. Powell, V. Marojevic, and G. Geraci,
``UAV-Assisted Attack Prevention, Detection, and Recovery of 5G Networks,''
IEEE Wireless Communications, vol. 27, no. 4, pp. 40--47,
2020.

\bibitem{VianaJ2022Access}
J. Viana, J. Madeira, Nidhi, P. Sebasti\~{a}o, F. Cercas, A. Mihovska, and R. Dinis,
``Increasing Reliability on UAV Fading Scenarios,''
IEEE Access, vol. 10, pp. 30959--30973,
2022.

\bibitem{Tyrovolas2022}
D. Tyrovolas, P. V. Mekikis, S. A. Tegos, P. D. Diamantoulakis, C. K. Liaskos, and G. K. Karagiannidis,
``On the Performance of HARQ in IoT Networking with UAV-mounted Reconfigurable Intelligent Surfaces,''
in 2022 IEEE 95th Vehicular Technology Conference (VTC2022-Spring), pp. 1--5,
2022.

%5
\bibitem{Song2021}
H. L. Song and Y. C. Ko,
``Beam Alignment for High-Speed UAV via Angle Prediction and Adaptive Beam Coverage,''
IEEE Transactions on Vehicular Technology, vol. 70, no. 10, pp. 10185--10192,
2021.


\bibitem{yuPPO2021}
C. Yu, A. Velu, E. Vinitsky, Y. Wang, A. Bayen, and Y. Wu,
``The surprising effectiveness of PPO in cooperative, multi-agent games,''
in Advances in Neural Information Processing Systems,
2021.


\bibitem{Chen2024}
C. Chen, Z. Guo, D. Luo, Y. Xu, and H. Duan,
``UAV Cooperative Air Combat Maneuvering Decision-Making using GRU-MAPPO,''
in 2024 IEEE 18th International Conference on Control \& Automation (ICCA),
2024.

%9


\bibitem{Lyu2023}
G. Lyu and M. Li,
``Multi-agent Cooperative Control in Neural MMO Environment Based on MAPPO Algorithm,''
in 2023 IEEE 5th International Conference on Artificial Intelligence Circuits and Systems (AICAS),
2023.


\bibitem{Huang2022}
H. Huang, T. Chen, H. Wang, R. Hu, W. Luo, and Z. Yao,
``MAPPO method based on attention behavior network,''
in 2022 10th International Conference on Information Systems and Computing Technology (ISCTech),
2022.

\bibitem{Cai2023}
W. Cai, X. Huang, Y. Chen, and Q. Guan,
``Multi-agent Cooperative Control in Neural MMO Environment Based on MAPPO Algorithm,''
in 2023 IEEE 5th International Conference on Artificial Intelligence Circuits and Systems (AICAS),
2023.

%12


\bibitem{Yan2024}
M. Yan, W. Guo, H. Zheng, and T. Qin,
``Joint NTP-MAPPO and SDN for Energy Trading Among Multi-Base-Station Microgrids,''
IEEE Internet of Things Journal, vol. 11, no. 10,
2024.

\bibitem{Peng2022}
R. Peng, S. Chen, and C. Xue,
``Collaborative Content Caching Algorithm for Large-Scale ISTNs Based on MAPPO,''
IEEE Wireless Communications Letters,
2022.

\bibitem{10021296}
W. Liu, B. Li, W. Xie, Y. Dai, and Z. Fei,
``Energy Efficient Computation Offloading in Aerial Edge Networks With Multi-Agent Cooperation,''
IEEE Transactions on Wireless Communications, vol. 22, no. 9, pp. 5725--5739,
2023.

\bibitem{li2023adaptive}
B. Li, W. Liu, W. Xie, N. Zhang, and Y. Zhang,
``Adaptive Digital Twin for UAV-Assisted Integrated Sensing, Communication, and Computation Networks,''
IEEE Transactions on Green Communications and Networking, vol. 7, no. 4, pp. 1996--2009,
2023.

\bibitem{wang2024uav}
W. Wang, X. Xu, M. Bilal, M. Khan, and Y. Xing,
``UAV-Assisted Content Caching for Human-Centric Consumer Applications in IoV,''
IEEE Transactions on Consumer Electronics, vol. 70, no. 1, pp. 927--938,
2024.


%17
\bibitem{Guan2023}
Y. Guan, S. Zou, K. Li, W. Ni, and B. Wu,
``MAPPO-Based Cooperative UAV Trajectory Design with Long-Range Emergency Communications in Disaster Areas,''
in 2023 IEEE 24th International Symposium on a World of Wireless, Mobile and Multimedia Networks (WoWMoM),
2023.





\bibitem{Jiang2024}
B. Jiang, J. Du, C. Jiang, Z. Han, and M. Debbah,
``Underwater Searching and Multiround Data Collection via AUV Swarms: An Energy-Efficient AoI-Aware MAPPO Approach,''
IEEE Internet of Things Journal, vol. 11, no. 7,
2024.


\bibitem{Kang2024}
J. Kang, J. Chen, M. Xu, Z. Xiong, Y. Jiao, L. Han, D. Niyato, Y. Tong, and S. Xie,
``UAV-Assisted Dynamic Avatar Task Migration for Vehicular Metaverse Services: A Multi-Agent Deep Reinforcement Learning Approach,''
IEEE/CAA Journal of Automatica Sinica, vol. 11, no. 2, pp. 430--445,
2024.


\bibitem{10279067}
Z. Lv, G. Niu, L. Xiao, C. Xing, and W. Xu,
``Reinforcement Learning Based UAV Swarm Communications Against Jamming,''
in ICC 2023 - IEEE International Conference on Communications, pp. 5204--5209,
2023.


\bibitem{10506576}
Y. Cai, H. Yao, Y. Gong, F. Wang, N. Zhang, and M. Guizani,
``Privacy-Driven Security-Aware Task Scheduling Mechanism for Space-Air-Ground Integrated Networks,''
IEEE Transactions on Network Science and Engineering, vol. 11, no. 5, pp. 4704--4718,
2024.

%%%%%/////

\bibitem{9475535}
S. Yin and F. R. Yu,
``Resource Allocation and Trajectory Design in UAV-Aided Cellular Networks Based on Multiagent Reinforcement Learning,''
IEEE Internet of Things Journal, vol. 9, no. 4, pp. 2933--2943,
2022.

\bibitem{9725258}
N. Zhao, Z. Ye, Y. Pei, Y.-C. Liang, and D. Niyato,
``Multi-Agent Deep Reinforcement Learning for Task Offloading in UAV-Assisted Mobile Edge Computing,''
IEEE Transactions on Wireless Communications, vol. 21, no. 9, pp. 6949--6960,
2022.

\bibitem{10720868}
M. Jia, L. Zhang, J. Wu, Q. Guo, G. Zhang, and X. Gu,
``Deep Multiagent Reinforcement Learning for Task Offloading and Resource Allocation in Satellite Edge Computing,''
IEEE Internet of Things Journal, vol. 12, no. 4, pp. 3832--3845,
2025.

%%%%%/////25


\bibitem{10624759}
L. Miuccio, S. Riolo, M. Bennis, and D. Panno,
``Design of a Feasible Wireless MAC Communication Protocol via Multi-Agent Reinforcement Learning,''
in 2024 IEEE International Conference on Machine Learning for Communication and Networking (ICMLCN), pp. 94--100,
2024.


\bibitem{Guan2023MAPPOBasedCU}
Y. Guan, S. Zou, K. Li, W. Ni, and B. Wu,
``MAPPO-Based Cooperative UAV Trajectory Design with Long-Range Emergency Communications in Disaster Areas,''
in 2023 IEEE 24th International Symposium on a World of Wireless, Mobile and Multimedia Networks (WoWMoM),
2023.


\bibitem{Kang2023CooperativeUR}
H. Kang, X. Chang, J. V. Misic, V. B. Mi\v{s}i\'{c}, J. Fan, and Y. Liu,
``Cooperative UAV Resource Allocation and Task Offloading in Hierarchical Aerial Computing Systems: A MAPPO-Based Approach,''
IEEE Internet of Things Journal, vol. 10, pp. 10497--10509,
2023.


\bibitem{Qin2024}
P. Qin, Y. Fu, J. Zhang, S. Geng, J. Liu, and X. Zhao,
``DRL-Based Resource Allocation and Trajectory Planning for NOMA-Enabled Multi-UAV Collaborative Caching 6G Network,''
IEEE Transactions on Vehicular Technology, vol. 73, no. 6, pp. 8750--8765,
2024.




%5////////////

\bibitem{Viana2024}
J. Viana, H. Farkhari, P. Sebasti\~{a}o, L. M. Campos, K. Koutlia, B. Bojovic, S. Lag\'{e}n, and R. Dinis,
``Deep Attention Recognition for Attack Identification in 5G UAV Scenarios: Novel Architecture and End-to-End Evaluation,''
IEEE Transactions on Vehicular Technology, vol. 73, no. 1, pp. 131--146,
2024.




\bibitem{SyntheticJViana}
J. Viana, H. Farkhari, P. Sebastiao, S. Lagen, K. Koutlia, B. Bojovic, and R. Dinis,
``A Synthetic Dataset for 5G UAV Attacks Based on Observable Network Parameters,''
arXiv preprint arXiv:2211.09706,
2022.


\bibitem{3GPP36777}
3GPP,
``Technical Specification Group Radio Access Network; Study on Enhanced LTE Support for Aerial Vehicles,''
\url{https://portal.3gpp.org/desktopmodules/Specifications/SpecificationDetails.aspx?specificationId=3231}.

\bibitem{3GPP38901}
3GPP,
``Technical Specification Group Radio Access Network; Study on channel model for frequencies from 0.5 to 100 GHz,''
\url{https://portal.3gpp.org/desktopmodules/Specifications/SpecificationDetails.aspx?specificationId=3173}.


%///////////////


\bibitem{nr_module}
N. Patriciello, S. Lagen, B. Bojovic, and L. Giupponi,
``An E2E simulator for 5G NR networks,''
Simulation Modelling Practice and Theory, vol. 96, pp. 101933,
2019.



\bibitem{Nr2024}
Centre Tecnol\`{o}gic de Telecomunicacions de Catalunya (CTTC),
``NR module Documentation,''
\url{https://cttc-lena.gitlab.io/nr/nrmodule.pdf},
Accessed: 04 September 2024,
2024.


\bibitem{viana2025pcafeatured}
J. Viana, H. Farkhari, P. Sebastiao, and V. P. Gil Jimenez,
``PCA-Featured Transformer for Jamming Detection in 5G UAV Networks,''
arXiv preprint arXiv:2412.15312,
2025.

\end{thebibliography}
\end{document}